\documentclass[namedreferences]{SolarPhysics}
\usepackage[optionalrh]{spr-sola-addons} 
\usepackage{graphicx}        
\usepackage{color}           
\usepackage{url}             

\newcommand{\ave}[1]{{\langle}#1{\rangle}}
\newcommand{\N}{{\textsf{N}}}
\newcommand{\Qz}{{\textsf{Q0}}}
\newcommand{\Qone}{{\textsf{Q1}}}
\newcommand{\Pz}{{\textsf{P0}}}
\newcommand{\Pone}{{\textsf{P1}}}
\newcommand{\Spot}{{\textsf{S}}}
\newcommand{\A}{{\textsf{A}}}
\newcommand{\B}{{\textsf{B}}}
\newcommand{\C}{{\textsf{C}}}
\newcommand{\D}{{\textsf{D}}}
\newcommand\arcsec{{\mbox{$^{\prime\prime}$}}}

\newcommand{\TSIv}{{\mbox{$\mathtt{TSI}_V$}}}
\newcommand{\TSIc}{{\mbox{$\mathtt{TSI}_C$}}}
\newcommand{\TSIr}{{\mbox{$\mathtt{TSI}_R$}}}
\newcommand{\deltaTSI}{{\mbox{$\delta\mathtt{TSI}$}}}
\newcommand{\Ir}{{\mbox{$I${\rm r}}}}
\newcommand{\FT}{{\rm FT}}
\newcommand{\QQ}{{\rm Q}}

\newcommand{\araa}{   {\it Ann. Rev. Astron. \&\ Astrophys.}}
\newcommand{\aap}{    {\it Astron. Astrophys.}}

\newcommand{\apj}{    {\it Astrophys. J.}}
\newcommand{\apjl}{   {\it Astrophys. J. Lett.}}
\newcommand{\apjs}{   {\it Astrophys. J. Suppl.}}

\newcommand{\solphys}{{\it Solar Phys.}}

\newcommand{\jmlr}{   {\it J. Mach. Learn. Res.}}
\newcommand{\msai}{   {\it Mem. Soc. Astron. Atal.}}

\begin{document}

\begin{article}

\begin{opening}

\title{Modeling Total Solar Irradiance Variations Using Automated Classification Software On Mount Wilson Data\\ {\it Solar Physics}}

\author{R.K.~\surname{Ulrich}$^{1}$\sep
        D.~\surname{Parker}$^{1}$\sep
        L.~\surname{Bertello}$^{1}$\sep
        J.~\surname{Boyden}$^{1}$
       }
\runningauthor{Ulrich, et al.}
\runningtitle{Modeling TSI Variations From MWO Data}

   \institute{$^{1}$ Department of Physics and Astronomy, Univerfsity of California, Los Angeles 90095
                     email: \url{ulrich@astro.ucla.edu}  email:  \url{darylgparker@gmail.com}
		     email: \url{bertello@astro.ucla.edu} email: \url{boyden@astro.ucla.edu}\\ 
             }

\begin{abstract}
We present the results using the AutoClass analysis application available at \inlinecite{2002Autoclass}
which is a Bayesian, finite mixture model classification system developed by \inlinecite{1995...kdd-95}.  We apply this system to Mount Wilson Solar Observatory (MWO) intensity and magnetogram images and classify individual pixels on the solar surface to calculate daily indices that are then correlated with total solar irradiance (TSI) to yield a set of regression coefficients.  This approach allows us to model the TSI with a correlation of better than 0.96 for the period 1996 to 2007.  These regression coefficients applied to classified pixels on the observed solar surface allow the construction of images of the Sun as it would be seen by TSI measuring instruments like the Solar Bolometric Imager recently flown by \inlinecite{2004ApJ...611L..57F}.  As a consequence of the very high correlation we achieve in reproducing the TSI record, our approach holds out the possibility of creating an on-going, accurate, independent estimate of TSI variations from ground-based observations which could be used to compare, and identify the sources of disagreement among, TSI observations from the various satellite instruments and to fill in gaps in the satellite record.  Further, our spatially-resolved images should assist in characterizing the particular solar surface regions associated with TSI variations.  Also, since the particular set of MWO data on which this analysis is based is available on a daily basis back to at least 1985, and on an intermittent basis before then, it will be possible to estimate the TSI emission due to identified solar surface features at several solar minima to constrain the role surface magnetic effects have on long-term trends in solar energy output.
\end{abstract}
\end{opening}
\section{Introduction}
     \label{S-Introduction} 

      Observations of solar surface magnetic and intensity data reveal a myriad of features which evolve on a wide range of spatial and temporal scales.  These features are in turn related to variations in solar radiative output both in specific wavelengths and in total solar irradiance (TSI) \cite{1994svsp.coll..217H}.  
One challenge to ascertaining the relationship between particular solar surface features and variations in solar radiative output is the very wealth of data available both from individual solar observations and from archives of individual solar observations which in some cases, such as those available at Mount Wilson Solar Observatory (MWO), cover many decades.  Traditionally, such features were identified manually and categorized by experts into familiar categories such as sunspots, faculae, quiet sun, etc., a process which inevitably produces considerable variation in how, and into which, feature categories regions from solar images are classified.  Recently, researchers have applied automated Bayesian statistical pattern recognition systems to identify and characterize solar features, using expert knowledge of the magnetic and radiative characteristics of the recognized types of regions to guide the software (\opencite{2002ApJ...568..396T}, \citeyear{1998ESASP.418..979T}, \opencite{1997ESASP.415..477P}).  Our approach has a similar goal but uses classification based on individual pixel properties instead of pattern recognition based on spatial configurations of groups of pixels.
      
      An automated classification or pattern recognition approach is particularly promising for analyzing the large database of solar observations taken with the 150-foot solar tower at MWO which has been in daily operation since 1912.  Since the mid-1980's, the observations at MWO have included daily magnetograms and intensity grams taken in the 5250.2\AA\ Fe {\sc i} and 5237.3\AA\ Cr {\sc ii} lines.  The observing system has been discussed in a number of papers (\opencite{1983SoPh...87..195H}, \opencite{1991SoPh..135..211U}, \citeyear{2002ApJS..139..259U}, \citeyear{2009SoPh..255...53U}) and utilizes a Babcock magnetograph system that scans an entrance aperture over the solar surface following a boustrophedonic pattern to build up a full disk image.  Because the magnetic and intensity images are captured simultaneously, each pixel in a magnetic or intensity image based on a particular spectral line corresponds exactly to the solar surface area covered by the corresponding image pixels in other lines.  This process permits simultaneous observation of the same part  of  the solar surface in different wavelengths.
      
      One product of the observations is the magnetic field strength for which we use the absolute value of the measurement at $\lambda5250$: $|B_{\lambda{\rm 5250}}|$.  We used only absolute magnetic field values (measured in gauss) for the magnetograms since, \it a priori\rm, the sign of the field did not seem likely to affect the intensity of radiation from that pixel.  For simplicity of notation we will use $|B|$ for this parameter since this is the only field we use in this study.  
It is important to recognize that magnetic field strength values are dependent on a variety of factors including the spectral line and spectral sampling as well as the spatial resolution.  The values given for the classifications discussed in this paper are those from the 150-foot solar tower synoptic program without correction to a $\lambda5233$ scale as discussed by \inlinecite{2009SoPh..255...53U}. A second product from the $\lambda5250$ and $\lambda5237$ lines is an intensity-like image which we have dubbed an intensity ratio-gram.  These consist of solar image arrays for which each pixel $i$ has a value $\Ir_i$ that is the ratio of the $\lambda5250$ intensity $I_{\lambda{\rm 5250}}$ in that pixel to the $\lambda5237$ intensity $I_{\lambda{\rm 5237}}$ in the same pixel:
\begin{eqnarray}
Ir_i&=&{10000\ \left(I_{\lambda{\rm 5250}}/I_{\lambda{\rm 5237}}\right)_i\over\left(I_{\lambda{\rm 5250}}/I_{\lambda{\rm 5237}}\right)_{\rm peak}}
\end{eqnarray}
where the intensity ratio-gram values are based on an arbitrary scale in which 10000 is the value of the peak of each image's probability distribution function for $Ir_i$.  Prior to use in the classification analysis, the $\Ir$ images have been adjusted to remove the dependence on center-to-limb angle $\rho$ by fitting to a polynomial in powers of $X=[1-\cos(\rho)]$ up to $X^4$ but with the term in $X^2$ omitted.  The fitting is done twice with intermediate steps 1) to identify and remove plages and spots using a histogram analysis and 2) to remove a small E/W gradient probably due to a telluric feature in the $\lambda5250$ line.  This flattening is carried out on each image prior to summing into the daily average image.  Also, to avoid some observational deficiencies near the solar limb, we have restricted our analysis to pixels having $\sin\rho<0.95$.
Because the magnetic sensitivities of the lines differ, the intensity ratio produces an image in which the usual surface features-spots, faculae, {\it etc} --- are visible.  The advantage to the intensity ratio-gram is that, by taking the ratio of the two intensities, short term variations in seeing and 
atmospheric conditions can be greatly reduced or removed, yielding an image which is superior in terms of solar surface feature resolution to an image in either wavelength alone.

\section{Use of AutoClass to Classify Solar Surface Features in MWO Magnetogram and Intensity Ratio-grams} 
      \label{S-general}      

\subsection{The AutoClass Software} 
  \label{S-text}
      
      AutoClass is a classification code developed by \inlinecite{1995...kdd-95} 
based on a classical finite mixture model utilizing Bayesian statistics for determining the optimal class to which a particular observation belongs.  The observed parameters are referred to as attributes and can be quite general.  Each observation is called an instance and is described by the attributes.  For us the attributes consist of two quantities: the absolute value of the magnetic field and the intensity ratio.  Additional observed quantities could be added if available but for our current effort we have retained just these two.  The instances are the individual pixels.  These instances (data pairs for a single pixel) are treated in isolation from all other attributes such as the image to which they belong.  We did explore the addition of the center-to-limb position as an additional parameter but found it did not improve the performance of the classification.  Although the fact that pixels belong to specific images in a time sequence does not enter into the classification analysis, we do utilize this information to understand the interrelations between the classes.  

The classification can be carried out in two different modes.  The first mode is a search mode in which it conducts a search for the optimal classification scheme based on posterior Bayesian statistics applied to a specific set of observed data.  The search mode can be (1) unconstrained or open ended, in which AutoClass determines both the optimal number of classes and the characteristics of those classes or (2) constrained or user directed, in which certain limits, such as the number of classes to be found, can be imposed on the search.  The output classification consists of a set of classes. Each class is defined by a distribution function giving the joint probability of occurance of the attributes for a class member instance.  The parameters defining the probability distribution functions (pdf's) for each attribute are the primary output from the search mode classification.  In our case the data used to find the optimal classification scheme consists of all the data pairs for pixels on a selected subset of solar images spanning the full solar cycle.  The second mode is a predict mode in which a user-directed classification is carried out by the application of the known classes to a new data set to determine the probability each data instance belongs to each of the classes.  
For the MWO data the attributes are specifically the absolute value of magnetic field $|B|$ and the 5250/5237 intensity ratio $\Ir$.
      
      When run in the open ended search mode, AutoClass decides both the optimal number of classes into which the data should be divided and the characteristics (in terms of observable attributes such as intensity ratio and magnetic field) of those classes.  The result of running AutoClass in the search mode on a data set is an instance by instance (here, pixel by pixel) classification in which each instance/pixel in the searched dataset is given a weighted probability assignment to a subset of the total classes found in the search mode.  For example, if the search mode found $J$ classes for the entire data set, then a particular instance/pixel $i$ with attributes/measurements $|B|_i$ and 
$\Ir_i$, would be given a set of probabilities $p_{ij}$ of belonging to each of the classes $j$.  In some cases, the most likely class will have a probability near unity while in other cases two or more classes may have a high probability.  For many pixels the probabilities for the least likely classes may essentially vanish.  There is thus no absolute assignment of an instance/pixel to a particular class.  This mode requires the determination of the number of classes $J$ as well as the parameters defining the probability distribution function (pdf) for each class so that a search mode run is computationally intensive.
      
      When run in the predict mode, AutoClass analyzes a new data set using the results of the search mode (the value of $J$ as well as the parameters defining the pdf for each class) to assign each instance/pixel in the new data set into one or more of the $J$ classes found in the search mode and to assign a probability to each class assigned to that pixel.  In this analysis, no new classes are introduced and AutoClass assumes that any pixel in the new dataset can be categorized into one or more of the $J$ classes found in the search mode.  The product of AutoClass as run in the predict mode consists of the assignment of class membership probabilities $p_{ij}$ to every pixel.

\subsection{Application of AutoClass to MWO Data} 
  \label{App_MWO}
To run AutoClass in the search mode using MWO images, we needed to use a reduced-size data set in order to achieve an acceptable running time.  For this purpose we created a dataset consisting of 12 daily image pairs (MWO magnetogram; intensity ratio-gram), using days spread over the period 1996 to 2008 --- {\it i.e.}, for each year in the time period, we selected one day and for that day used a magnetogram and intensity ratio-gram for that day.  In order to reduce noise, we (1) used images consisting of the average of the magnetograms and the average of the intensity ratio-grams for each selected day, (2) eliminated days for which there was only 1 image and (3) selected days for which there were at least 8 pairs of magnetic and intensity images.  (The number of available images per day and type varies from 0 to 16 or more depending on weather and time of year.)  When assembled, this dataset contained measurements for 484,000 pixels from 12 pairs of magnetogram and intensity ratio-gram images over 12 years, each pixel having a measured value for the absolute magnetic field and the intensity ratio.  

As a ground-based observatory at a single site, the MWO system must cope with a variety of difficulties -- weather-induced gaps, sky transparency variations and image blurring known as seeing.  Apart from the observations where the sky transparency variations impacted the data, the variations of atmospheric seeing and other instrumental effects did not degrade the image quality since the entrance aperture has a size of 12\arcsec\ squared and the seeing image fluctuations are generally only 2 to 4 \arcsec\ on days of poor seeing.  The effect of sky transparency variations is felt primarily through their introduction of tracking errors into the system controlling the position of the solar image.  At a low level some unsatisfactory images have been left in the data set.  Individual daily classification data points have been examined when their results appear discordant and the tracking error mechanism appeared to have an impact on the data quality.  Since these points tend to occur on days when one or two observations were made, the easiest way to reduce their impact on the analysis was to restrict the data points to those days on which three or more observations were made.  
The most adverse months are January to April.  For some years, the end of December also tends to have poor data but this happened less than was the case for the later 4 months.  We return to this discussion below in the context of a detected discrepancy having a period of a year.  In a further note about systematic effects related to the observing system we point out that prior to winter 1993 the spectrograph experienced strong spectrograph seeing during periods of cold weather with the result that the magnetic fields were noiser at those times.  None of the presently used data is subject to this problem. 

A core problem is to determine the number of classes required to represent solar surface features.  AutoClass can make this determination based on its Bayesian statistical model of the data pairs.  However, our observational parameters come from a system where the spatial resolution does not isolate solar surface features that are uniform in their characteristics.  Consequently, it is probable that some classes are a result of a variable filling factor.  For example if $(|B|_{\FT},Ir_{\FT})$ and $(|B|_\QQ,Ir_\QQ)$ are the observational parameters respectively in a flux tube and the quiet sun and a particular solar surface pixel has $(|B|,\Ir)= f(|B|_{\FT},\Ir_{\FT})+(1-f)(|B|_\QQ,\Ir_\QQ)$, AutoClass may create separate classes for each of a limited range of $f$ between 0 and 1 whereas in fact there are only two classes present.  In the event that AutoClass finds excess classes from this mechanism, the extra classes will form a subspace where the $|B|,\Ir$ pairs are linearly dependent and the regression analysis will yield large, nearly cancelling coefficients.

	AutoClass was first run in the search mode with the number of classes to be used either fixed or left as a free parameter to be determined by the code.  The initial AutoClass search results yielded 37 classes.  The regression analysis discussed in the following section indicated that the filling factor mechanism might in fact be playing a role in the MWO data set.  Consequently we restricted the AutoClass search to have a maximum of 18 classes.  Although the ability of the 37-class classification to reproduce the TSI was slightly better than that for 18 classes, we believe that this result is not physically significant.  Furthermore, the 18-class set is easier to relate to known solar features.

Using the search mode results for the 18-class set, we then ran AutoClass in the predict mode on the daily average magnetograms and intensity ratio-grams for the period from May, 1996 to July, 2008.  For purposes of running AutoClass in predict mode, we modified the operation sequence so that the images for each day $n$ were treated as a single dataset containing a $(|B|,\Ir)$ measurement for each pixel. We indicate that pixel $i$ is found in image $n$ with the notation $i\in n$. Designating the total number of satisfactory days we analyze as $N$, the current data set has $N=3111$.  The predict mode run takes $(|B|_i,Ir_i)$ for each $i\in n$ and returns a set of the 10 most probable classes $p_{ij}$ for these pixels.  The restriction on the number of returned class probabilities per instance is a feature of the AutoClass package necessitated by the potentially large number of classes that AutoClass can identify. We were not able to modify this feature.  Because this set of probabilities is truncated, the sum of the probabilities of the returned classes does not reach unity and a renormalization of the class probabilities was necessary in order to ensure that the area index defined below properly represents the total solar surface.  

      The next step was to use the new datasets created by AutoClass in the predict mode to create for day $n$ an index for each class $j$:
\[ A_{jn}= (I_n)^{-1}\left(\sum_{i,i\in n}^{I_n} p_{ij}\right) \qquad {j=0\dots 17} \]
where $I_n$ is the number of pixels in image $n$.  Thus, the derived index $A_{jn}$ is the expected fractional area of the solar surface covered by class $j$ on day $n$.  Also, for each $i\in n$ we retain the position $x_i,y_i$ as the distance from the image center in the EW and NS directions relative to the image radius.  We use these positions to reconstruct a 2 dimensional image showing the locations of the different class pixels.  These images help in the identification of the classes with solar surface features of recognized type such as sunspots, plages, faculae and chromospheric network.

\section{Use of AutoClass Indices to Model TSI} 
      \label{S-features}      

It is important to observe that, so far in the process, there has been no introduction of any variables other than: (1) the measured intensity ratio and magnetic field values from the MWO images and (2) the AutoClass class probabilities and daily class indices, the latter of which were based solely on the observed number distribution functions of the data pairs $(|B|,\Ir)$ with each pixel being treated as an independent entity.  The creation of the daily class index from the sum over pixels associated with each image is a separate step outside the AutoClass system.  These indices are the fractions of the area of the apparent solar disk covered by each class and having the property:
\begin{eqnarray}
\hspace{1.1in}\sum_{j=0}^{J-1}A_{jn}&=&1\hspace{0.6in}(n=1\cdots 3111)\ .
\end{eqnarray}
We are also interested in averages over all the daily indices and use a bracket notation 
\begin{eqnarray}
\ave{A_j}&=&N^{-1}\sum_{n=1}^N A_{jn}
\end{eqnarray} 
to indicate such an average with a similar notation for other quantities.

In order to determine whether the daily indices $A_{jn}$ could be related to satellite TSI observations and solar surface features, we next performed a linear multiple regression to fit the 18 daily MWO/AutoClass indices for 3111 days between May, 1996 and July, 2008 to 3 daily TSI measurements for the same period from: (1) \TSIv, obtained from  
the VIRGO experiment on the cooperative ESA/NASA mission SoHO, \cite{Virgo2008}
(2) \TSIc, the composite VIRGO dataset (version d40) obtained from PMOD/\-WRC, Davos, Switzerland, unpublished data from the VIRGO Experiment on the cooperative ESA/NASA Mission SoHO (the creation of which is described in \opencite{2000SSRv...94...15F}); and (3) \TSIr\ the space absolute radiometric reference (SARR) \cite{1995AdSpR..16...17C,2004SoPh..224..209D}.  In this paper we illustrate the method using the \TSIv\ case while a subsequent paper will discuss the application of our methods to all three TSI time series as well as UV indicators.  The above symbols will be used whenever we refer to the time series data.

The regression fit we use contains only an overall constant and does not associate a constant with each class since a class with zero area must contribute nothing to the TSI.  Thus we seek coefficients $s_j$ giving a model $S$ for the virgo TSI, \TSIv, of the form:
\begin{eqnarray}
S_n&=& c_0 + \sum_{j=0}^{J-1} s_j\,A_{jn}
\end{eqnarray}
using the method of least squares.  Each of the coefficients $s_j$ corresponds to the TSI value the sun would produce in case it were covered entirely by a surface of class $j$.  
We can evaluate $c_0$ by averaging over all the $n$ daily cases:
\begin{eqnarray}
\ave{S} &=& c_0 + \sum_{j=0}^{J-1} s_j\,\ave{A_{j}}\ .
\end{eqnarray}
Since the $s_j$ have the value of $S$ that would prevail if $A_j=1$ and $s_j\ave{A_{j}}$ is the average contribution of class $j$ to $S$, we see that $s_j\ave{A_{j}}$ must sum to $\ave{S}$ and the value of $c_0$ must be zero.

We can use the fact that
the area indicies must sum to unity to introduce an arbitrary offset since 
any arbitrary constant $C$ subtracted from the $\TSIv$ would result in an equal constant subtracted from all the coefficients $s_j$:
\begin{eqnarray}
S_n-C&=&\sum_{j=0}^{J-1} s_j\,A_{jn}-C\sum_{j=0}^{J-1} A_{jn} = \sum_{j=0}^{J-1}(s_j-C)A_{jn}
\end{eqnarray}
where the requirement of all areas summing to unity is used in the first form of the right hand side of this equation.

Thus the inclusion of a constant just provides a zero point offset for fitting parameters and is not constrained in the fit.  If we drop one class from the model equation, then a non-zero constant will have value equal to the $\TSIv$ for the dropped class.  We do not use this approach because we find that all the derived values of $s_j$ depend on which class is dropped.  To avoid introducing a dependence on which class is dropped from the fit, we have adopted a bootstrap method applied to the solutions without a constant as described below.

For most classes, the values of $s_j$ are close to the average TSI so we adopt the arbitrary zero point offset to be the global average of the $\TSIv$ and consider deviations:
\begin{eqnarray}
\delta s_j&=&s_j - \ave{S}\label{equna}\\
\noalign{\smallskip}
\noalign{\leftline{and}}
\noalign{\smallskip}
\delta S_n&=&S_n - \ave{S}
=\sum\limits_{j=0}^{J-1} \delta s_j\,A_{jn}\ .
\end{eqnarray}
For our current investigation concentrating on the VIRGO data the difference between the observation and simulation for day $n$ is then:
\begin{eqnarray}
\deltaTSI_n &=& \TSIv_n - S_n\ .
\end{eqnarray}  
We have chosen to define $\deltaTSI_n$ with the above a sign to simplify our discussion of the time dependence of $\deltaTSI$ near solar minimum in Section \ref{SMS}.  

We find that the regression analysis is sensitive to the details of the data set to which the regression is applied.  This may be due to the constraint on the sum of the $A_{jn}$ indicies \cite{1996MaxEntEcon} or due to the near linear dependence of the observed data pairs brought about by the filling factor effect.  Although the quality of the fit is consistent when small changes are made to the fitted data, the coefficients themselves vary.  We determine the most representative values of the coefficients from a bootstrap approach:
\begin{enumerate}
\item The base pool of data consists of the actual set of 3111 daily observations each of which is the 18 class indices combined with the \TSIv\ for that day.
\item A modified data set is formed by a method called randomly selecting with replacement from the full base pool 3111 times.
 The modified data set contains some daily observations more than once and omits others.  The temporal order of the
daily observations is not preserved in the modified data set.  However, including the \TSIv\ as part of the daily observation prevents the fitting coefficients from being impacted by this temporal scrambling.
\item The fitting coefficients are computed from the modified pool and recorded.
\item Steps 2 and 3 are repeated up to 100,000 times to create distribution functions for each of the classes.
\item The median from each distribution function is adopted as the appropriate value for that class.
\end{enumerate}
This method allows us to estimate the uncertainty in the fitting coefficients.  As an additional measure of the uncertainty in the derived fitting coefficients we applied the method of dropping one class for a 17-class fit.  Since the constant in that case represents the \TSIv\ for that class, we subtracted the global average \TSIv\ from that constant so it could be treated the same way as the remaining classes.  Also, to avoid special focus on any particular class, we carried out this analysis for all of the classes to obtain 18 values for the fitting coefficient for each of the 18 classes.  We use the average of coefficients from these 18 cases as an estimate we term the 17-class fit coefficients.
We then compared the average of the coefficients for the 18 cases to the coefficients found from the bootstrap method.  For all classes the coefficients derived in these two ways agreed with a difference much smaller than the error indicated from the bootstrap method.  The worst cases were for classes 13, 17, 15 and 16 where the bootstrap coefficients are 
$-0.53,\;-324,\;54$ and $20.9$ W$\;$m$^{-2}$ while the 17-class fit coefficients are $-4.58,\;-327,\;56$ and $19.3$ W$\;$m$^{-2}$ respectively.

 \begin{figure}
\begin{center}
\parbox{4.5in}{\begin{center}
\resizebox{4.5in}{!}{\includegraphics{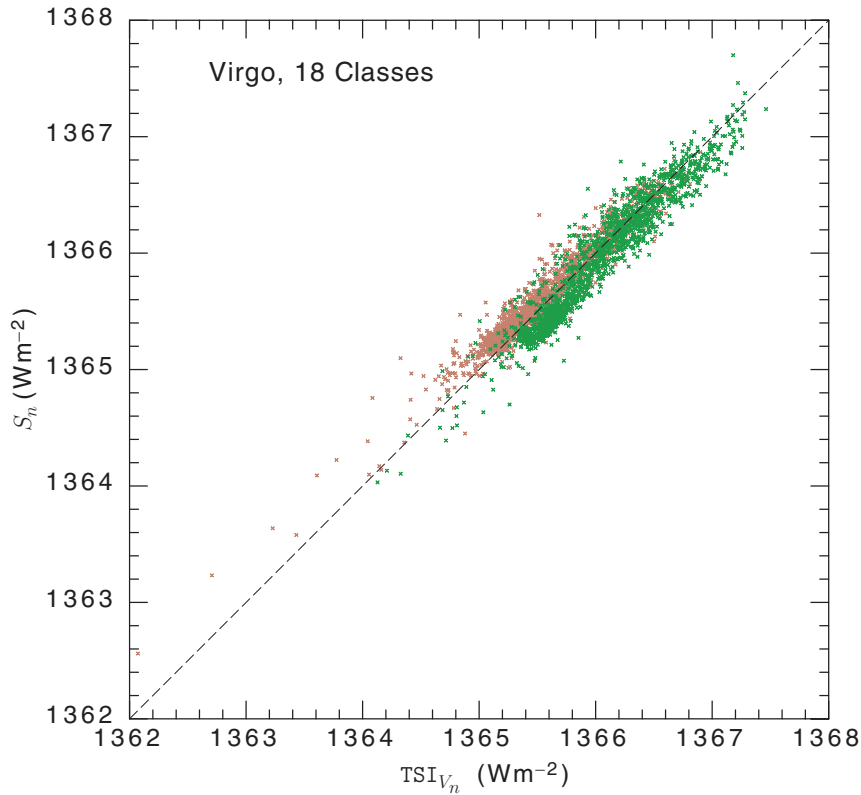}}
\end{center}
}
\caption{The relationship between observed and modeled values of the TSI during the period 1996 to 2008.  The period analyzed coincides with solar cycle 23.  During this interval, the short term variations are well followed but there is also a long-term trend for which the model deviates from the observations.  To make this effect apparent, we have divided the period into a first half colored brown and a second half colored green in order to illustrate the smaller dispersion over the shorter periods.}
\label{fig1}
\end{center}
\end{figure}

\begin{figure}
\begin{center}
\parbox{4.8in}{\begin{center}
\resizebox{4.8in}{!}{\includegraphics{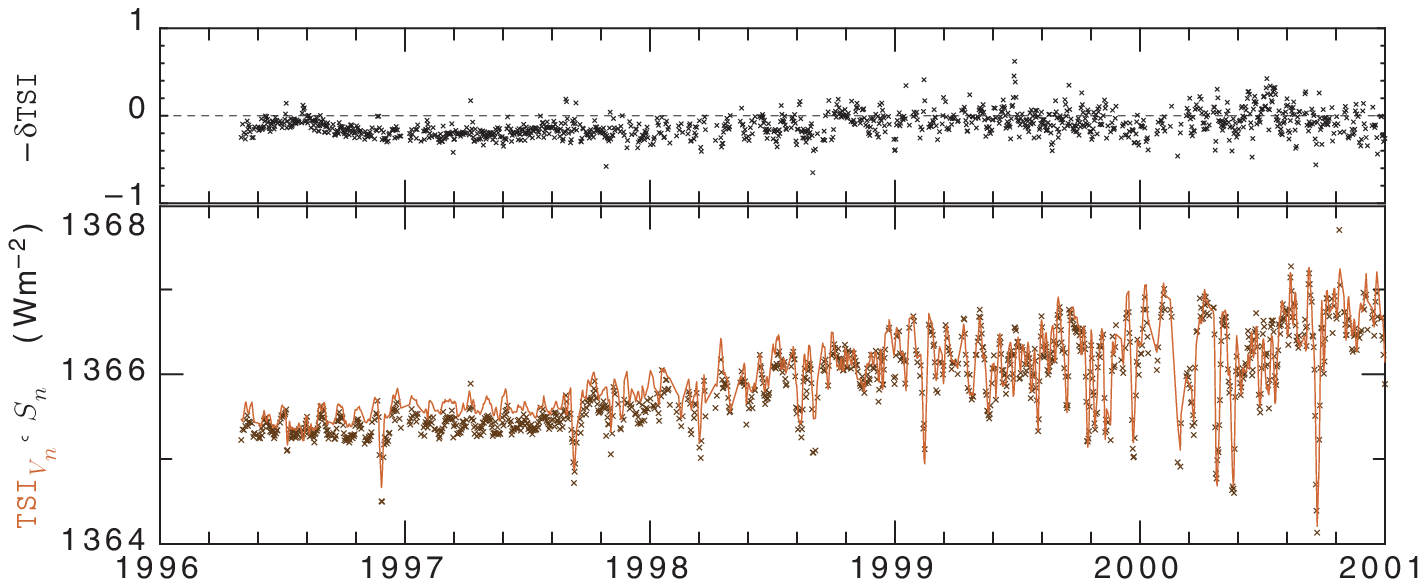}}
\resizebox{4.8in}{!}{\includegraphics{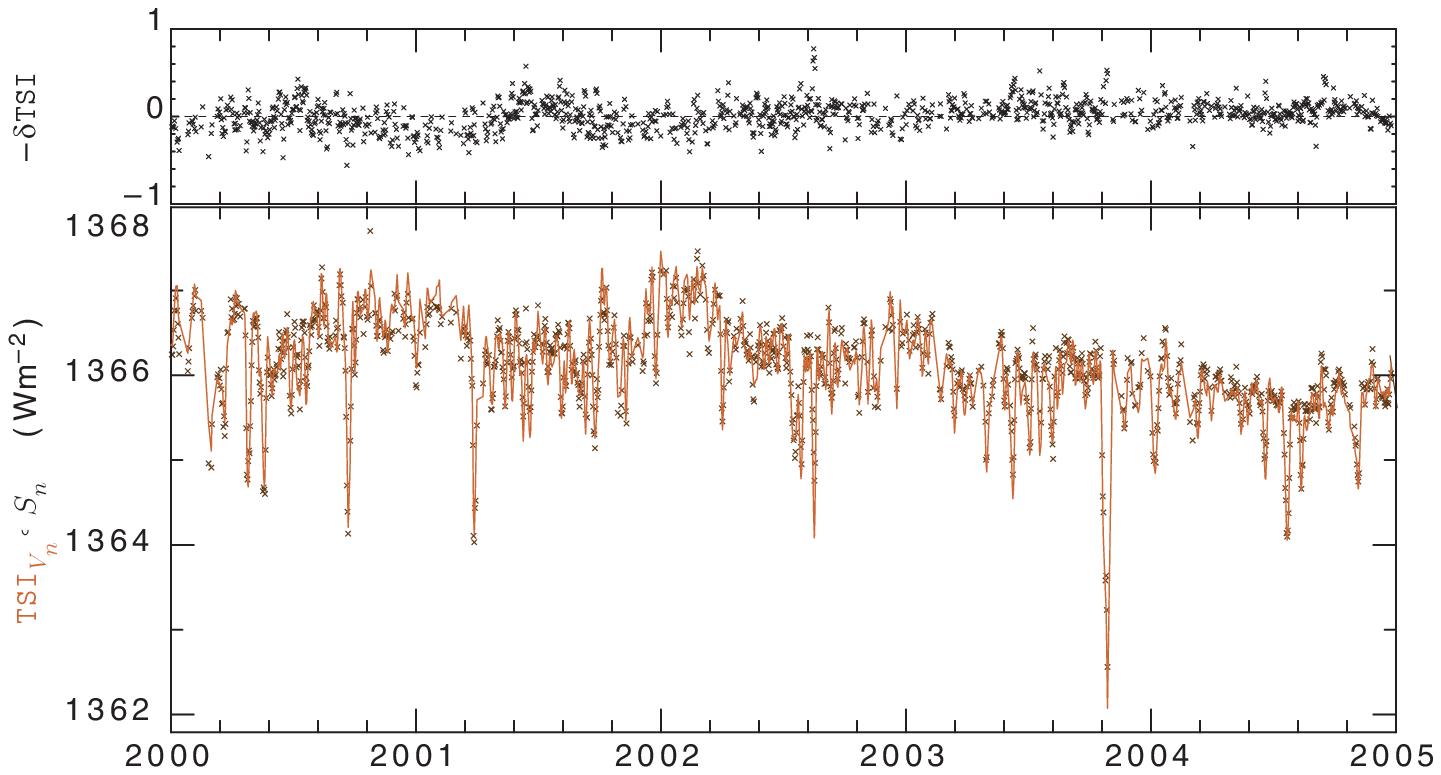}}
\resizebox{4.8in}{!}{\includegraphics{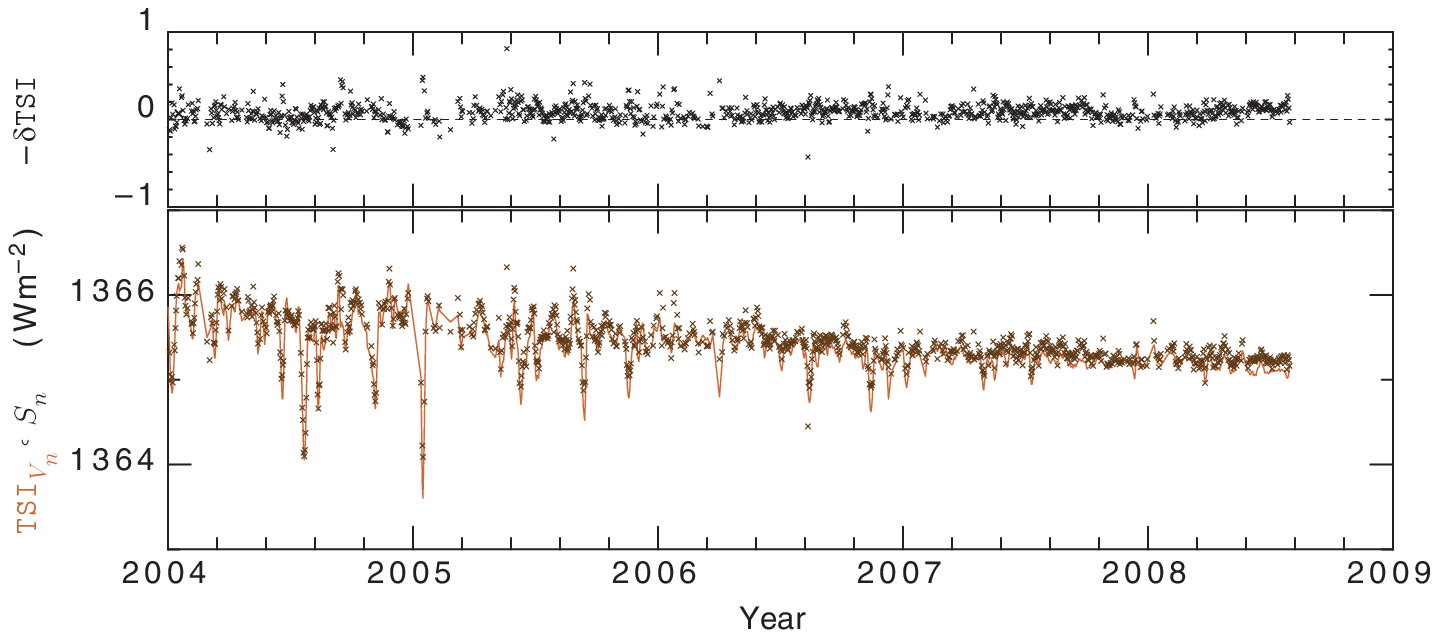}}
\end{center}
}
\caption{This figure composed of three sets of two time series comparing the observed VIRGO TSI: $\TSIv_n$ (orange solid lines) to the simulation, $S_n$ (the crosses).  The bottom of each pair shows the values of the observed and simulated TSI's while the top of each pair shows the deviation: simulated minus observed TSI values (note that $\deltaTSI$ is defined as observed minus simulated).  The full time series has been divided into three sets of 5 years with an overlap of one year between successive sets so that patterns can be seen near the dividing points.}
\label{fig2}
\end{center}
\end{figure} 
\begin{figure}
\begin{center}
\parbox{4.5in}{\begin{center}
\resizebox{4.5in}{!}{\includegraphics{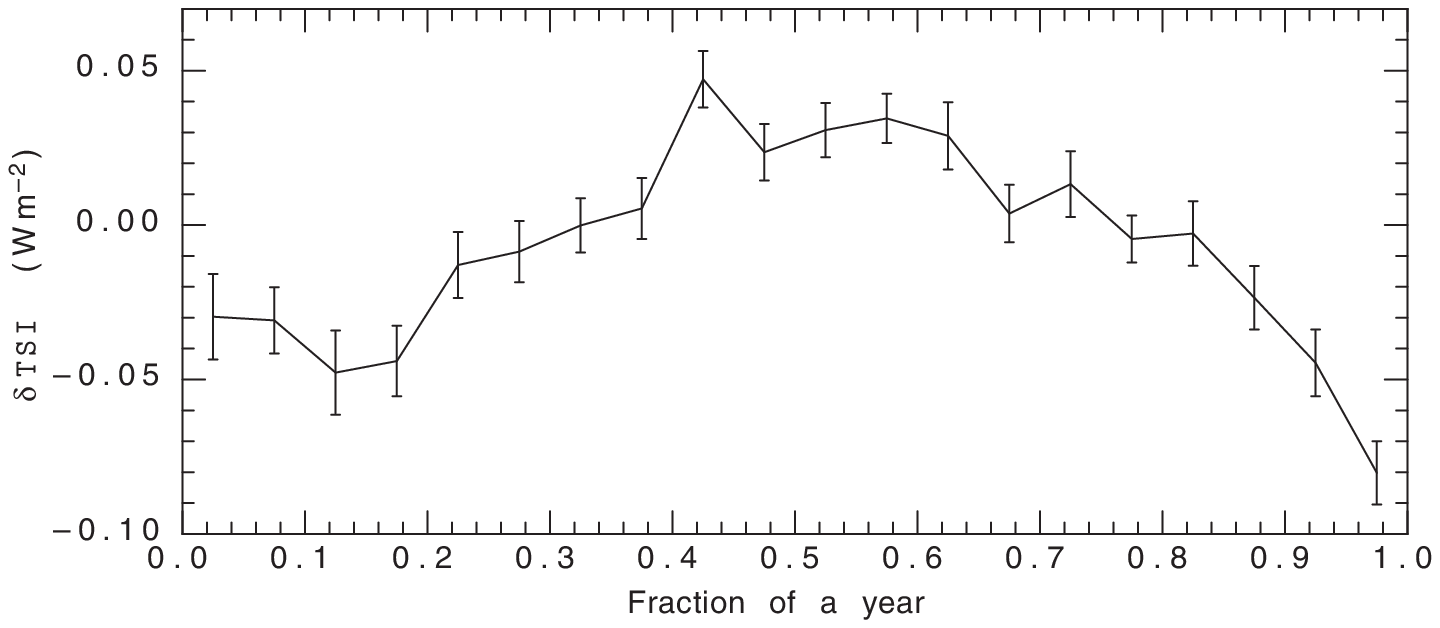}}
\end{center}
}
\caption{The annual pattern of deviation in the simulation of the VIRGO TSI relative to the observations.  Each point is the result of binning the $\deltaTSI$ values according to the fraction of a year past the beginning of each year.  For the full data set, each bin which is 1/20th of a year has between 87 points in February and 214 points in July.  The error bars shown are the errors of the mean for each bin.  Prior to binning, the time series was linearly detrended so that the long-term trend would not increase the estimate of the errors of the mean.
}
\label{fig3}
\end{center}
\end{figure}
    
      The correlation between the MWO daily class indices and the VIRGO TSI measurement for the 3111 day period was  0.9565 for a fit to the 18 classes while a fit with the full 37 classes yielded 0.9625.  Figure \ref{fig1} showes the scatter plots for the MWO model, $S$, and the VIRGO TSI, \TSIv\ measurements.  The tightness of the relationship between the model and observations is evident.  In fact it is somewhat better than is indicated by the full set of cases since there is a long-term trend of the model relative to the observations.  The color of the plotted points has been chosen so as to highlight this effect with the early part of the decade being green and the ending part of the decade being orange. The cause for this offset is not known since there have been no changes to the MWO instrumentaton that could give such a systematic drift. Figure \ref{fig2} showes comparative time series for the MWO fitted simulations and the observed VIRGO TSI data.  If we average over the first 6-year period and compare it to the last 6-year period we find that the value of $S$ has decreased by only $0.07\,$W$\,\!$m$^{-2}$ whereas the observed VIRGO TSI has decreased by $0.31\,$W$\,\!$m$^{-2}$ for the same periods.  The spectrographic system at MWO has been held constant over these periods.  

The deviation plots show patterns with a period of a year.  This is illustrated in Figure \ref{fig3} which shows the result of a superposed epoch analysis with a one year period starting at the first of the year.  We do not have an explanation for this pattern.  We also see the systematic trend of the observed points to fall relative to the simulation for the full time period.  The simulation is essentially the same for the two minima.  There is a possible effect related to the tilt angle of the magnetic field vector in the polar regions.  However, the symmetry between north and south regions leads to the expectation that such an effect would have a dominant component with a period of 6 months.  Occasional bad observations that are not eliminated by our quality control procedures tend to have larger than appropriate magnetic fields and their associated model TSI values tend to be high.  Since such images are more common during the winter and spring months, this effect would produce a result out of phase compared to what is shown in Figure \ref{fig3}.
      
      The datasets created by running AutoClass in the predict mode can also be used to create resolved images of the sun as it would be seen by a telescope which detected TSI directly.  Thus, because each dataset created in the predict mode contains for each pixel the probabilities for the 10 most probable classes for that pixel, and since each predicted class has a fit coefficient derived from the multiple regression of the daily MWO indices against the various TSI measurements, it is possible to create a solar TSI image by assigning to each pixel a value equal to the sum of the per class coefficients times the class probabilities plus the constant derived from the multiple regression.  Images of this sort are discussed below for groups of classes rather than for individual classes.

\section{Class Groups}  
     \label{S-groups}
\subsection{Cross-Correlation Analysis}
The solar surface is covered by a number of recognized types of feature that typically include quiet sun, chromospheric network, plages/faculae and sunspots.  The 18 (or 37) types of solar surface identified by AutoClass clearly makes a finer distinction of solar surface features than is currently recognized.  In addition, we are concerned that the filling factor effect continues to generate artificial classes.  AutoClass itself provides no guidance beyond the grouping of pixels into the classes.  We utilize the additional information provided by the time series of daily indices to search for groupings among the classes.  Thus if we find several classes that have highly correlated daily indices we can determine how this group as a whole might relate to the recognized solar surface features.  Furthermore, the values of $|B|_j$ and $\Ir_j$ determined by AutoClass are a significant indicator of the nature of each class.

Our grouping of the classes is accomplished through the study of the cross-correlation matrix formed from the 18 time series of the daily class indices.  AutoClass creates a class number according to its internal processing and generally gives low class numbers to those classes having the largest number of members.  However, the class numbers themselves have no apparent useful information.  The procedure we use to find class groups is to identify a subset of the classes that can serve as root classes for their group.  Based on a tentative identification of a class as a potential root class, we can arrange all the remaining classes in order of decreasing cross-correlation coefficient.  All classes having a cross-correlation coefficient above some threshold are then identified as being members of that group.  The root class for the next group can be taken as the class having the highest cross-correlation coefficient below the threshhold.  This process can be repeated until all classes are members of some group.  In some cases, the group has only one member if its root class has no other classes with a cross-correlation coefficient above the threshold.  This stepwise grouping process is simple and yielded good results. More
complex problems might benefit from the global approach detailed in \inlinecite{2008JMLR...9..485B},
which decomposes the covariance into groups using a 
maximum-likelihood criterion based partly on the inverse
covariance matrix where the group structure is directly encoded.

\begin{table}
\caption{Class groups and their cross correlations}\label{tab1}
\parbox{\hsize}{Each cell represents the correlation coefficient between the two classes at the ends of the row and column intersecting at the cell.  The lines mark out the classes that have been identified as groups with the group symbol either left or above the classes making up the group.  The correlations between classes within each group are higher than with classes outside the group.}
\smallskip

\parbox{\hsize}{
\resizebox{\hsize}{!}{\includegraphics{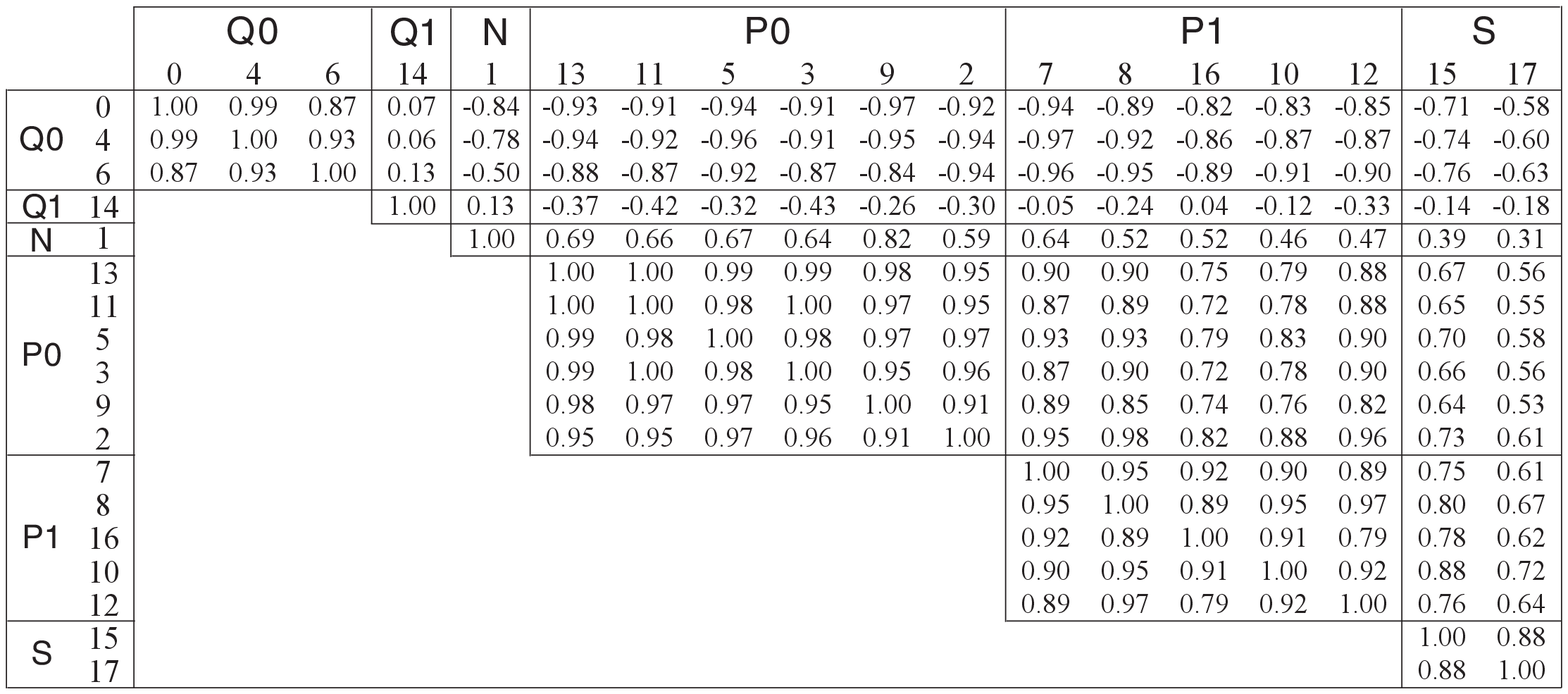}}}
\end{table}

The lowest number class found by AutoClass has the largest number of pixels and examination of the values of $|B|$ and $\Ir$ strongly suggest that this class corresponds to the quiet sun.  As long as we follow the above procedure of beginning each group by taking the root class to be the class with the highest cross correlation below the threshhold from the previous group, the sequence of groups will represent a systematic increase in the level of activity.  In fact the values of the $|B|_j$ and $Ir_j$ determined by AutoClass for the classes within these groups confirms this expectation.  The value of the threshhold cross-correlation coefficient to define the dividing point between successive groups could not be held constant for all the groups.  This process allows us to find six groups of classes that we name Quiet 0 (\Qz), Quiet 1 (\Qone), Network (\N), Plage 0 (\Pz), Plage 1 (\Pone) and Spot (\Spot).  

The result of the above procedure is shown in Table \ref{tab1}.  For most of the groups the assigned classes are clearly distinctly associated with that group and there is little flexibility in making the assignment.  However, for the two plage groups \Pz\ and \Pone, the association is less clear.  For example, if the root class of group \Pz\ were chosen as class C5 instead of class C13 then class C12 would have been assigned to \Pz\ instead of \Pone.  However, based on the magnetic field strength, it is more appropriate in group \Pone.
The number of classes within each of the above groups is 3, 1, 1, 6, 5 and 2 respectively.  The fact that the Plage groups have the largest number of classes as members is probably a consequence of the filling factor mechanism.

\subsection{Class Group Properties}
\label{Group_prop}

\begin{table}
\caption{Properties of class groups}  
\parbox{\hsize}{This table gives the principal properties of the classes and groups.  The classification parameters $|B|_j$ and $Ir_j$ are in gauss and are $10000\times$(intensity ratio) respectively.  The $\delta s_j$ are from the bootstrap/regression analysis of the TSI Virgo observations and are given in W$\,\!$m$^{-2}$ deviations from the average of the TSI as given by Equation \ref{equna}.  The center-to-limb distances of the classes and groups are shown here using the averages $\left<r_j\right>$ of $p_{ij}r=p_{ij}\sin(\rho)$ over all pixels.}\label{tab2}
\smallskip
\parbox{\hsize}{
\resizebox{\hsize}{!}{\includegraphics{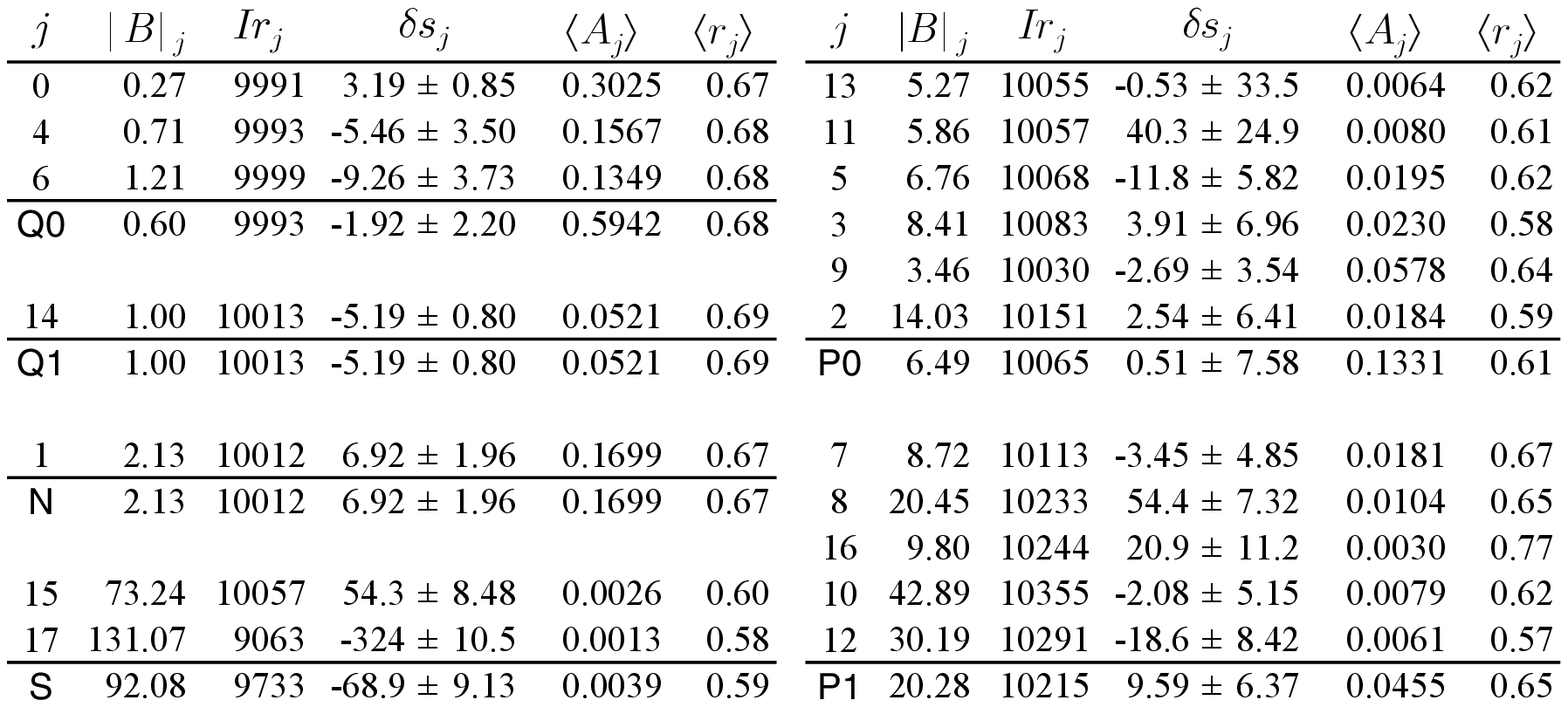}}}
\end{table}
\begin{figure}
\begin{center}
\parbox{4.8in}{\begin{center}
\resizebox{4.8in}{!}{\includegraphics{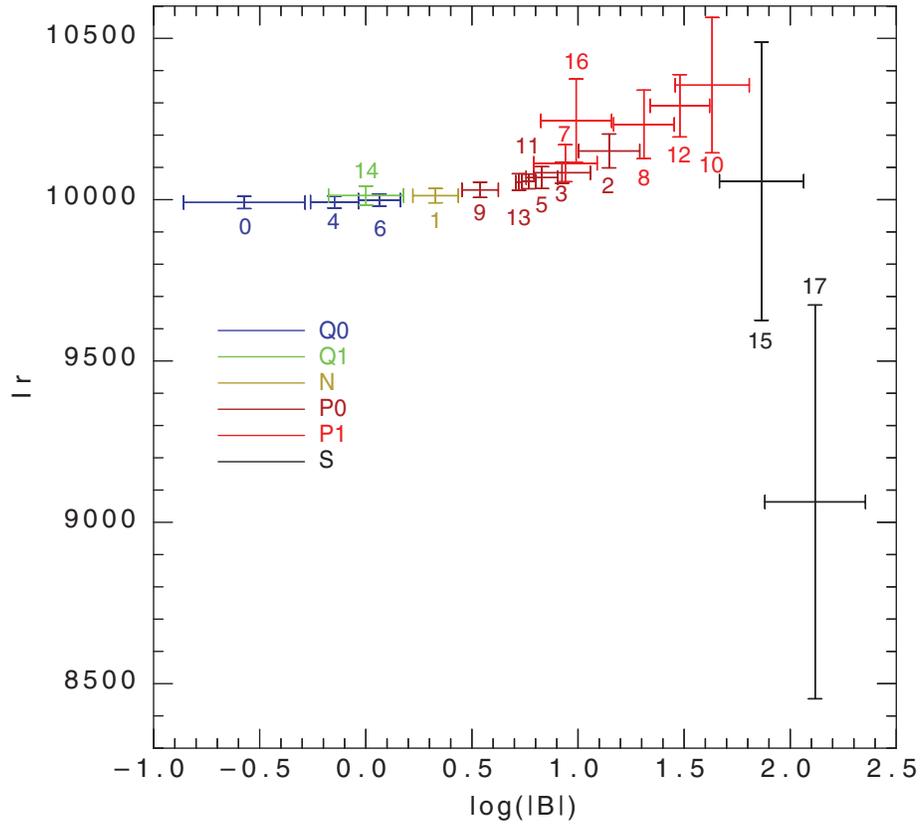}}
\end{center}
}
\caption{This figure shows the values for $|B|_j$ and $Ir_j$ along with the Gaussian widths determined by AutoClass for the 18 classes.  In order to have a uniform spacing of the $|B|_j$ values, they are plotted in a logarithmic format.  For these the Gaussian width is shown as $0.4343\, \delta|B|/|B|$.  The membership of each class in its assigned group is indicated by the color code.  For several of the classes in group \Pz, the properties overlap and are not clearly displayed.  This is consistent with our conclusion that some of the classes are in fact the consequence of the filling factor effect.
}
\label{fig4}
\end{center}
\end{figure}

The relationship of the classes and groups to solar properties is of great interest and is shown in Figure \ref{fig4} and Table \ref{tab2}.  Although the classes in the groups are similar, their regression coefficients are often of different sign within the groups suggesting that there may be more parameters than needed for the basic fit.  However, the uncertainties derived from the bootstrap analysis indicate that sign of the coefficient, especially for the \Pz\ and \Pone\ groups, is not significant.  The concentration of the classes with the least certain fitting coefficients in these two groups is supportive of the hypothesis that uncertainty in the fit comes from the flux-tube/filling factor mechanism discussed in Section~\ref{App_MWO}.  The systematic increase in magnetic field strength in going through the sequence (\Qz, \Qone, \N, \Pz, \Pone, \Spot) is evident.  The intensity ratio is an indicator of the brightness of the upper photosphere compared to the low photosphere with larger values indicating a brighter upper photosphere/chromosphere zone.  Its dependence on the magnetic field strength is mostly along a well defined curve where the upper atmosphere is brighter where the field is stronger.   Exceptions occur within the \Pz\ and \Pone\ groups where the dependence is erratic. In the case of the \Qone\ group (a single class) the chromosphere is brighter than other portions of the solar surface with similar field strength.  The average areas given in the last column for each group are averaged over the full solar cycle and show that over 80\%\ of the solar surface is covered by the \Qz, \Qone\ and \N\ groups.  Nonetheless, these groups have low cross-correlation coefficients and represent distinct components of the solar surface.

A useful diagnostic for the classes and groups is their distribution over the solar surface.  We select only those pixels whose probability of membership in a single class $j$ is over 90\%\ and utilize their center-to-limb angles $\rho_{ij}$ as measured by $r_{ij}=\sin(\rho_{ij})$.  For each image $n$ we calculate the average of $p_{ij}r_{ij}$ to get a measure of the average pixel distance $(r_n)_{\rm ave\, \mathit{n}}=(p_{ij}r_{ij})_{\rm ave\, \mathit{n}}/A_n$ where the notation $(a)_{\rm ave\, \mathit{n}}$ denotes an average over the pixels in image $n$.  Since we are interested in the distribution of the class pixels relative to the analyzed image, we rescale the values of $r_{ij}$ by dividing by 0.95.  The daily averages are then averaged over the full data set to yield $\ave{r_j}$. We have also considered the {\it rms} spread in $r_{ij}$ denoted by $\Delta r_j$.  For reference we note that a quantity distributed uniformly over the solar image would have $\ave{r_j}=2/3$, $\ave{\Delta r_j} = \left(1/2 - 4/9\right)^{1/2} = 0.236$.  This analysis clarifies the distinction between class 16 and the other \Pone\ and \Pz\ classes with similar magnetic field strength -- class 16 is systematically nearer the limb than the others and has a higher value of $\Ir$ as a consequence.

\begin{figure}
\begin{center}
\parbox{\hsize}{\begin{center}
\resizebox{\hsize}{!}{\includegraphics{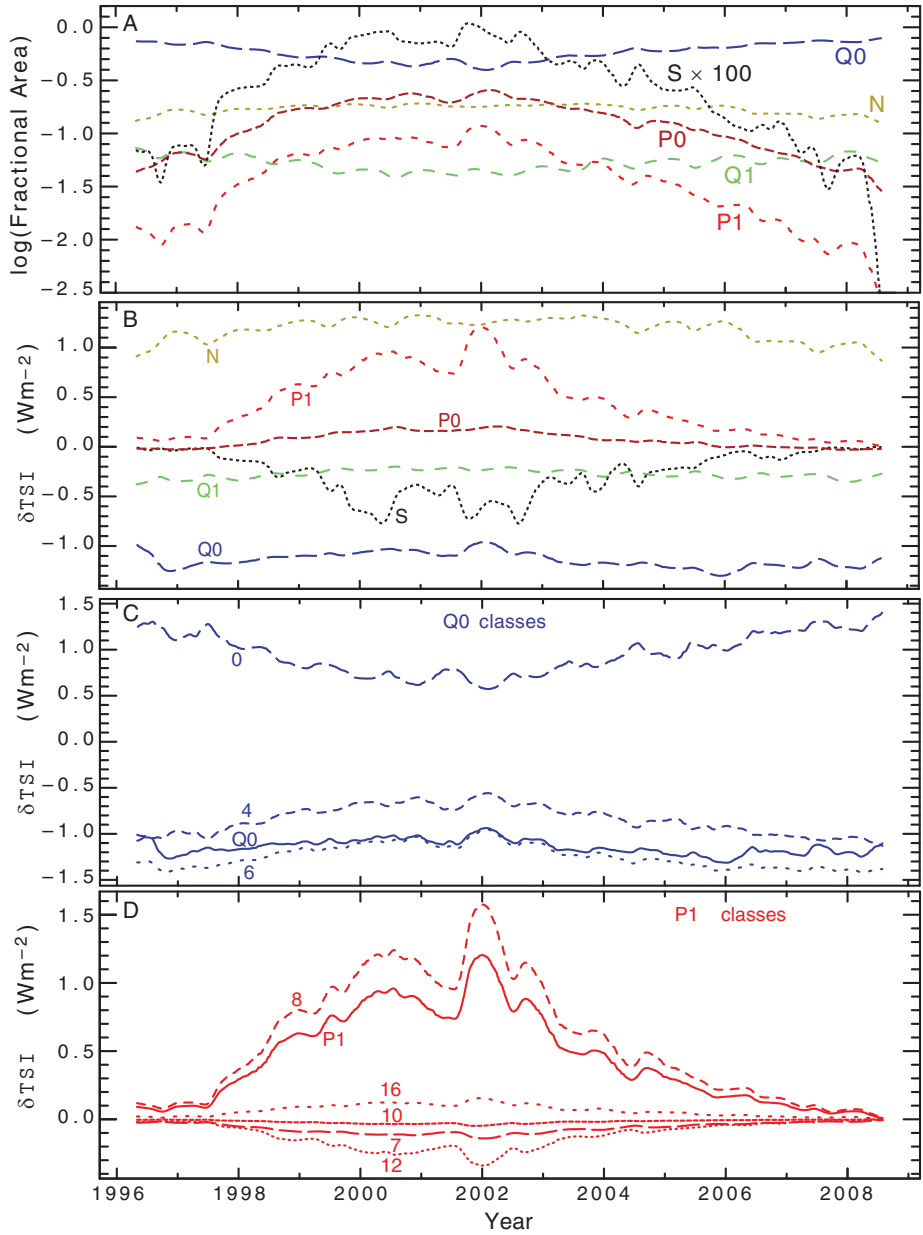}}
\end{center}
}
\caption{The smoothed time trends of the classes and class groups.  The time dependences have been smoothed with a Gaussian with a width of 30 observations.  The top figure, panel \A, shows the areas, the second panel \B\ shows the contributions to the TSI.  In both cases we show the classes grouped according to the system outlined above.  The third and fourth panels \C\ and \D\ show the components of two of the groups: that for the \Qz\ group in panel \C\ and that for the \Pone\ group in panel \D.}
\label{fig5}
\end{center}
\end{figure}

\begin{figure}
\begin{center}
\parbox{\hsize}{\begin{center}
\resizebox{\hsize}{!}{\includegraphics{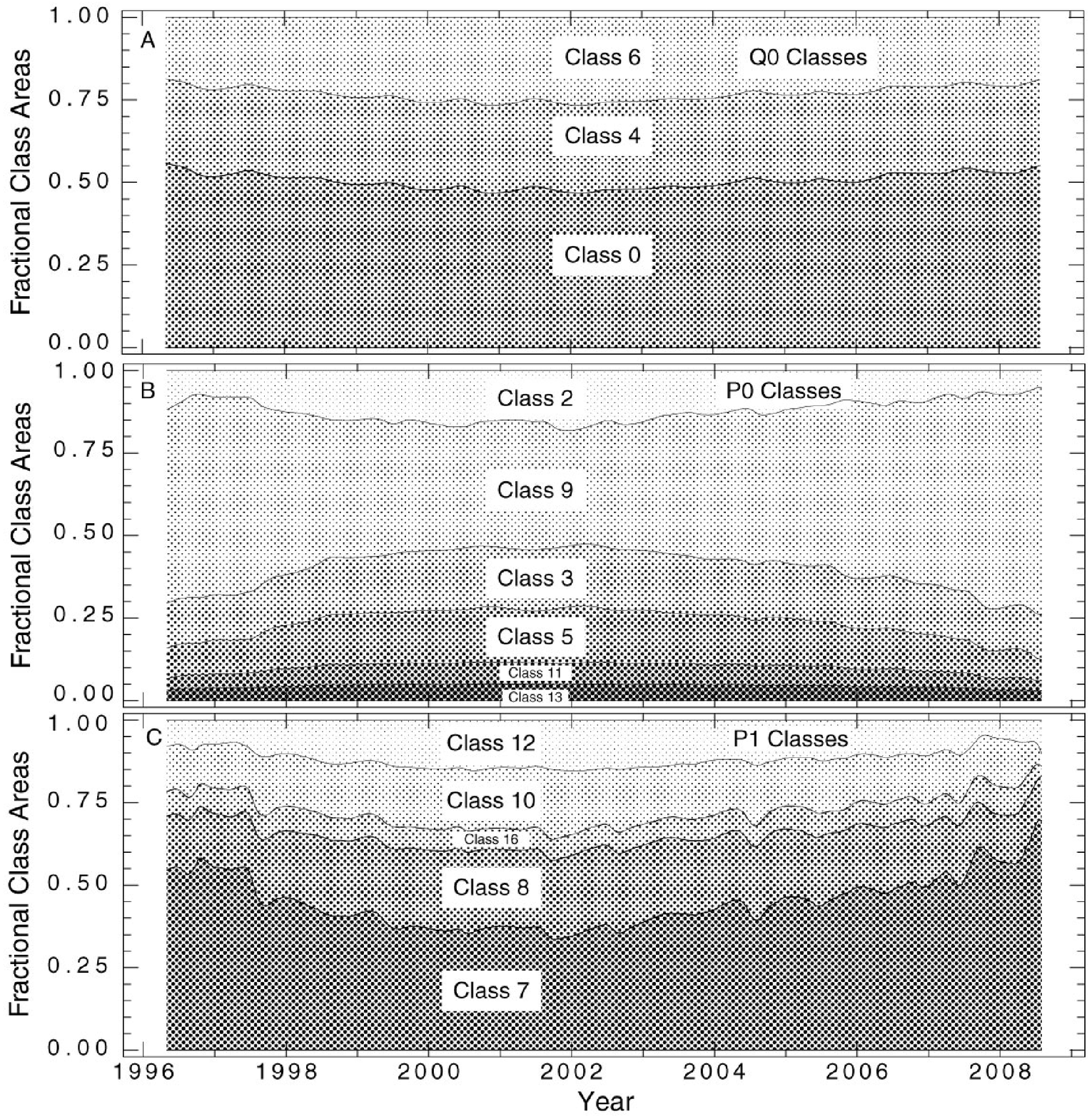}}
\end{center}
}
\caption{The smoothed time trends of the fractional area the classes within a class group occupy relative to the total area of that class group.}
  \label{fig6}
\end{center}
\end{figure}

Much of the variability dominating the cross-correlation coefficients comes from the rotational modulation of the solar surface coverage.  Changes due to the 11-year cycle are also of interest.  The growth or decline of the fractional coverage of a particular class or group of classes may provide a clue as to the underlying process of the solar dynamo.  The relative sizes of the coefficients $\delta S_j$ allows the groups \Pone\ and \Spot\ to have a major impact on the TSI even though the areas themselves may be small.  We reduce the rotational modulation in the areas and TSI model by smoothing all parameters with a Gaussian function having a width of 30 days.  We show in Figures \ref{fig5} and \ref{fig6} respectively the smoothed trends for the contributions to the TSI and the changing breakdown of the class areas within three groups of classes.  Several of the groups influence the TSI but show only a restricted dependence on the solar cycle: \Qz, \Qone\ and \N. The case of the \Pone\ group is interesting because the \textsf{D} panel of Figure \ref{fig5} suggests that class C8 accounts for most of the TSI variation.  
However, all classes within the \Pone\ group contribute a nearly constant fractional area to the \Pone\ group and their contributions to TSI differ primarily because of the differing values of their $s_j$ coefficients.  The substantial cancellation effects amoung the $s_j$ coefficients within the \Pone\ group are typical of an underdetermined system having more parameters than constraints.  The least square fitting exploits the small differences between the class time dependencies through these cancellations.

Because the \Qz\ group occupies a major fraction of the available surface area, when the \Pz\ and \Pone\ groups increase their area of coverage, the \Qz\ and \Qone\ groups decrease their area because they provide the areas supplying the \Pz\ and \Pone\ groups.  During the rising portion of the solar cycle, the conversion of the quiet group areas into plage group areas is associated with a rising TSI.  Consequently some of the increase in the TSI is attributed to a negative coefficient for the \Qz\ and \Qone\ groups.

\section{Images}

\begin{figure}
\begin{center}
\parbox{\hsize}{\begin{center}
\resizebox{\hsize}{!}{\includegraphics{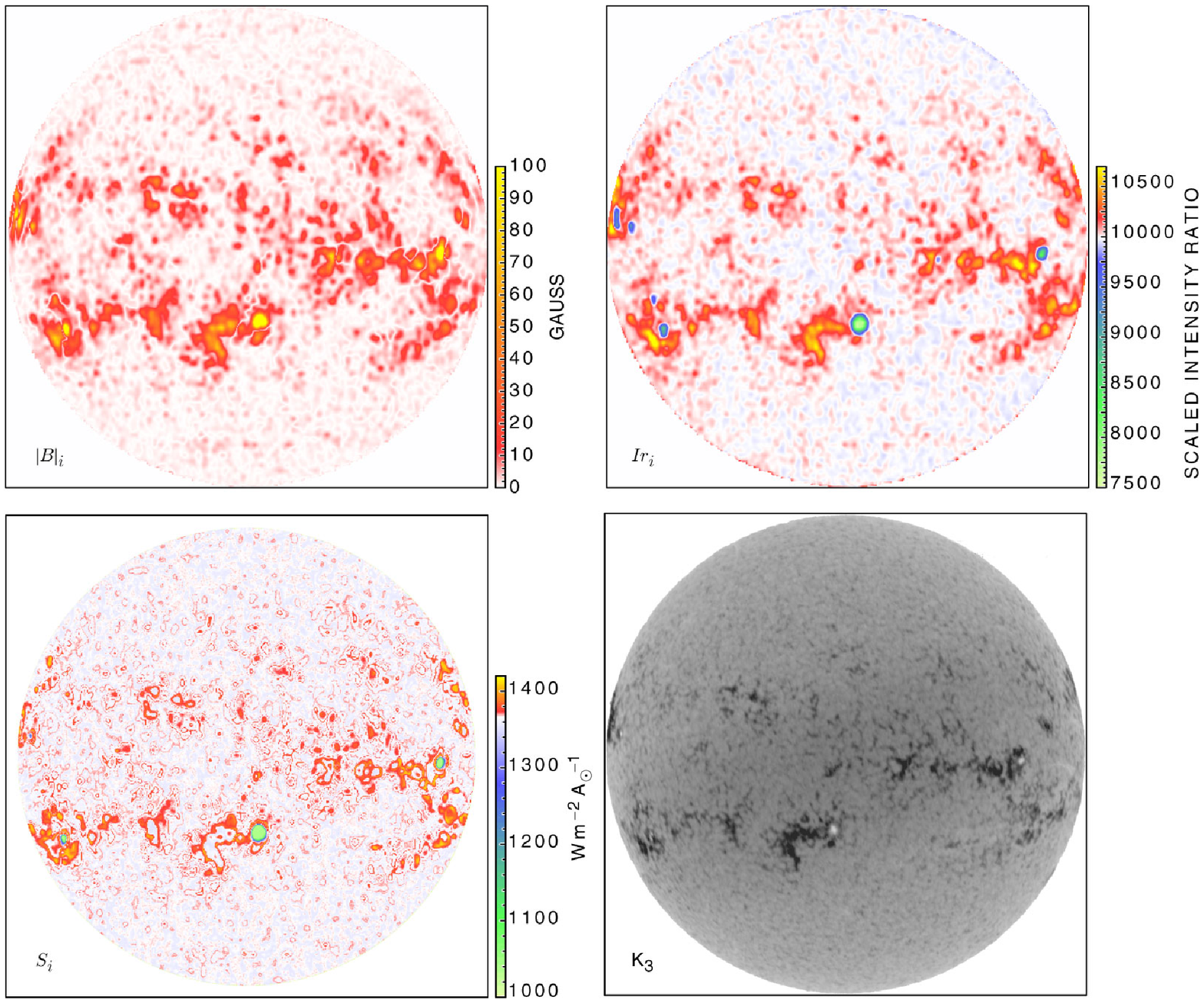}}
\end{center}
}
\caption{The absolute magnetic field and intensity ratio data images along the top row and the derived model TSI image on the lower left. The lower right image is from the Meudon synoptic spectroheliogram program and shows the intensity in the $K_3$ line of CaII and is shown for comparison to more traditional measures of solar magnetic activity.  The MWO and Meudon images are for May 16, 2002 but are not temporally synchronized so that magnetic contours from MWO do not match this CaII $K_3$ image although the temporal changes from MWO to Meudon are not large.
}
\label{fig7}
\end{center}
\end{figure}
\begin{figure}
\begin{center}
\parbox{\hsize}{\begin{center}
\resizebox{\hsize}{!}{\includegraphics{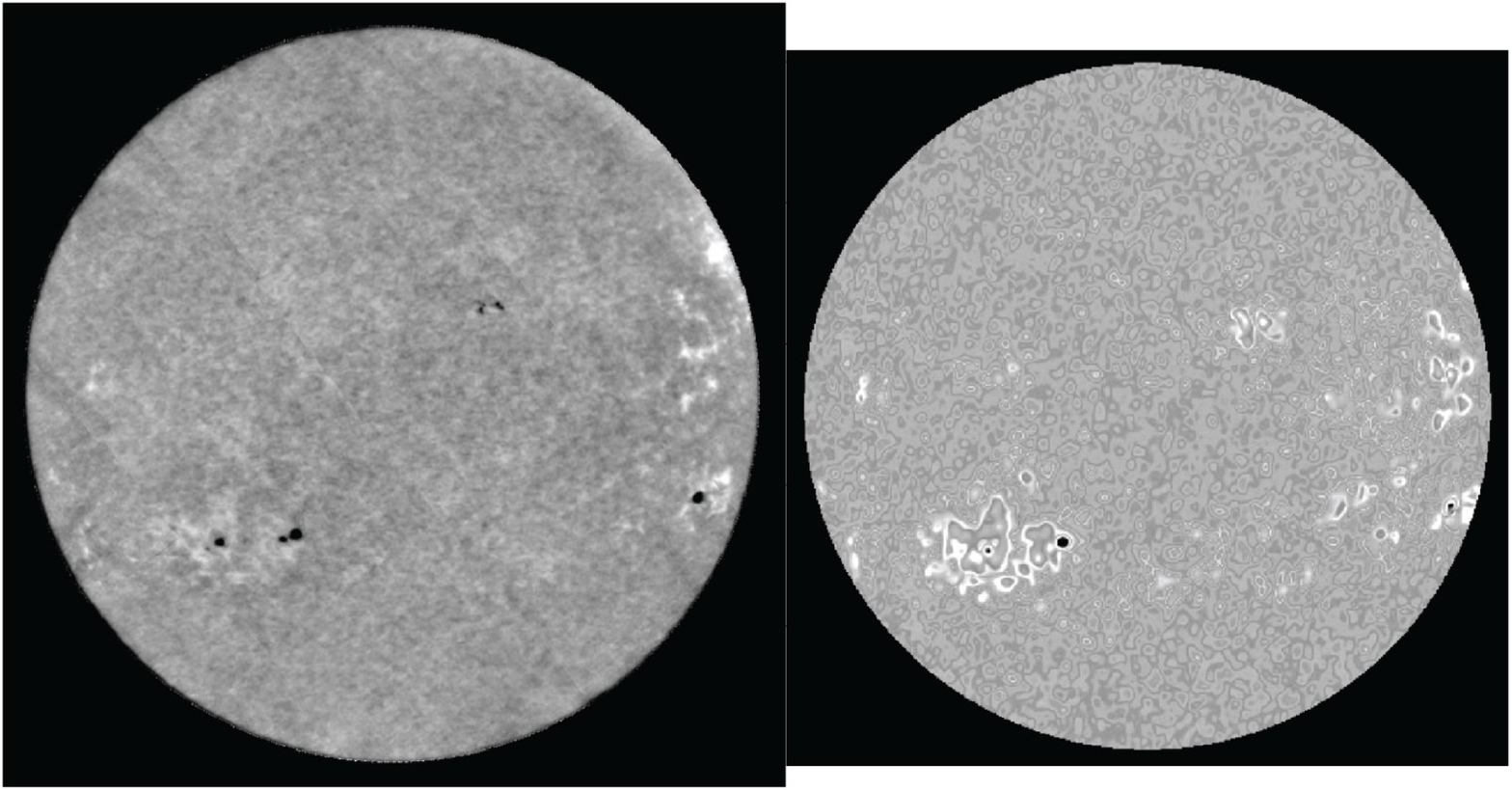}}
\end{center}
}
\caption{This figure compares the spatially resolved image from the Solar Bolometric Imager (SBI) obtained on 9 Sep.\ 2003 by \protect\inlinecite{2004ApJ...611L..57F}
to a model TSI image prepared according to the procedures described here.  Both images are on a greylevel scale where black is 95\%\ and white is 103.8\%\ of the average.  The SBI data has been corrected for limb darkening and was provided to us by the authors in a digital format so that we could create the image with the grey levels matching ours to illustrate the key differences compared to our TSI image.  Note that the images are shown with the same angular scale and that the SBI image goes to the solar limb whereas our image only goes to a point 5\%\ in from the solar limb and has a smaller image.  
}
\label{fig7a}
\end{center}
\end{figure}

The relationship between the observable parameters and the derived change in the model TSI has both expected and unexpected parts.  In this section we mostly show a number of representations of the solar image for a representative day of 16 May 2002.  We also use data obtained on 9 Sep.\ 2003 with one image from MWO and another image from the balloon-borne Solar Bolometric Imager (SBI) instrument \cite{2004ApJ...611L..57F}.  16 May 2002 has a moderate level of activity and not a high level of activity wherein the regions might cover more of the solar surface and tend to merge or interact.  It is not unusual in any obvious way and is of good quality data not impacted by terrestrial sky transparency variations.  Figure \ref{fig7} shows images of the absolute value of the magnetic field and the intensity ratio along the top row.  For comparison on the lower right is a familiar representation of the chromospheric structure as measured by a Ca K$_3$ spectroheliogram adapted from the synoptic program carried out at the Paris Observatory at Meudon.  This K$_3$ image is displayed here as a negative in order to be more consistent with the magnetic and intensity ratio images that have been created entirely digitally.  The close correspondence between magnetic field strength and the K$_3$ intensity is well known.  The intensity ratio has not been used before as a diagnostic of solar surface features.  However, the image shown in Figure \ref{fig7} is very similar to that for $|B|$ with two differences: 1) the strong magnetic field regions have a reduced value of the intensity ratio, and 2) in regions of very weak magnetic field, the intensity ratio continues to become lower even though the field itself is lower by only a very small amount.  When these features are classified and converted to a contribution to the TSI, a map of the solar surface can be created.  The result for the sample day is shown in the lower left.  Note that the positive contributions to the TSI are not as widely distributed as are the magnetic and intensity ratio features.  

The balloon-borne Solar Bolometric Imager (SBI) has provided us with a real image of the sun of the sort that our algorithm attempts to produce.  The comparison between the real SBI image and our model is shown in Figure \ref{fig7a}.  The image based on the class coverage fractions is not the local photometric contrast but instead is the local contribution to the change in the TSI.  The SBI image shows the local contrast wherein the deviations near the limb are enhanced relative to ours.  The model TSI image differs from the observed in the important characteristic that it has placed an excessive fraction of the irradiance deviation in a bright rim around the plage rather than leaving it distributed somewhat uniformly in the plage or facular regions.  This pattern is most likely a consequence of the ring effect discussed above in Section \ref{Group_prop} and is related to the ill-determined coefficients in the plage groups.

\begin{figure}
\begin{center}
\parbox{\hsize}{\begin{center}
\resizebox{\hsize}{!}{\includegraphics{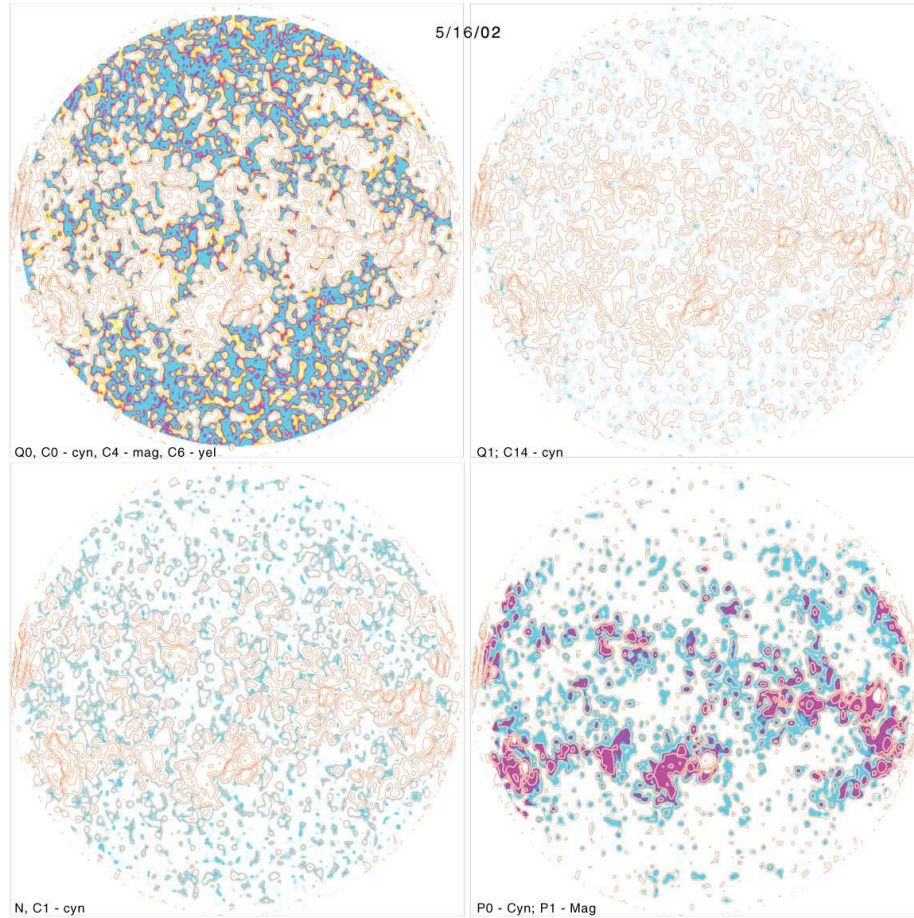}}
\end{center}
}
\caption{Images for 16 May 2002 formed from the classes using a CMY encoding.  In each image a class or a group of classes is assigned one of the three colors so that for the upper left image representing the breakdown of the \Qz\ group, class C0 is assigned the cyan color, class C4 is assigned the magenta color and class C6 is assigned the yellow color.  Each color is fully saturated when the probability of the pixel belonging to that class is 100\%\ and is white when the probability is 0\%.  Multiple colors are combined when the probability is shared among several classes.  The contours shown in orange on the figures are lines of constant $|B|$ starting with 2G then 7G, 20G, 70G, 200G and 700G. }
\label{fig8}
\end{center}
\end{figure}

The grouping of the classifications in Section \ref{Group_prop} is based exclusively on the cross correlations between the time series created using the fact that the pixels are from solar images.  The names of the groups imply an association with known solar structures although some of the groups are subdivisions of these previously known structures.  The sequence (\Qz, \Qone, \N, \Pz, \Pone, \Spot) is ordered in terms of increasing magnetic field.  The values of the field in each case is consistent with the fields for these structures except perhaps for the \N\ group (the single class C1) and the \Pz\ group where the fields are weaker than expected.  An important diagnostic of the nature of these groups comes from the distribution over solar images and the spatial proximity of pixels of each class or group to other classes and groups.  

We have created diagnostic images from the class groups using a CMY (Cyan, Magenta, Yellow) encoding wherein a one to three color {\tt tiff} image has each of the color strengths based on one or more class group probabilities -- each pixel is fully the assigned color if its probability of membership in the assigned class is 100\%.  Pixels having a significant probability of being in two or three classes displayed in the figure have their colors composed of those class value and can take on an appropriate intermediate hue.  Class group images created this way are shown in Figure \ref{fig8} for \Qz, \Qone, \N\ and (\Pz, \Pone).  The three classes making up the \Qz\ group are shown using the full CMY encoding, the plage groups are both shown in one figure using two colors while the remaining two groups have only one class and are show separately using the cyan color alone.  The \Spot\ group is not shown in a panel because the areas involved are very limited and can be seen adequately as the blue and green areas on the $Ir_i$ panel of Figure \ref{fig7}.  In order to aid in the coordination of these class group figures with the input magnetic field pattern, an overlay of contours of constant $|B|$ strength values is plotted on top of the class group plots.  

The \Pone\ magenta areas are clearly very concentrated about the high magnetic field positions while the \Qz\ areas clearly avoid the magnetic contours.  The gradation of the classes within the \Qz\ group is clearly a sequence with the quietest C0 class forming the center of the areas and being surrounded by the C4 and C6 classes in succession.  The group identified as \N\ is widely distributed as expected for the solar chromospheric network justifying our naming of this group.  The \Qone\ group appears to occur preferentially near the solar limb and this is probably the reason it is not well correlated with the other classes that have been grouped as \Qz. 

Referring back to Figure \ref{fig7}, it is evident that the plage areas have an alternating ring pattern where the central portions of the plage are not bright but are surrounded by a rather bright ring.  This pattern is not seen in either the $|B|_i$ or the $Ir_i$ figures.  According to Table \ref{tab2}, the \Pone\ group has both positive and negative $s_j$ coefficients with the area weighted average being positive.  In Figure \ref{fig7} the $S_i$ panel is computed from the full set of classes rather than the class groups.  In the parts of the figure with strong active regions characterized by sunspots or concentrations of \Pone\ areas, this alternating ring pattern is seen quite clearly.  In more isolated areas which only have \Pz\ or \Pone\ areas, the alternating ring pattern is not present.  This pattern comes primarily from the way the $s_j$ coefficients have been constrained in the fit to the TSI.  The most strongly positive coefficient is that for class 8 while the stronger magnetic field classes 10 to 12 have small or negative coefficients.  An examination of the spatial distribution of the class 8 pixels shows that they are found both associated with the plage regions and in areas well away from the stronger plage zones.  Evidently the widely-distributed pixels are essential for the fit to the TSI and lead to a high $s_j$.  Apparently this large coefficient yields a higher-than-needed contribution from the stronger plage regions so that an additional contribution from the class 10 to 12 pixels is not needed to match the TSI.  The close spatial proximity of the class 8 pixels and the class 10 to 12 pixels makes this cancellation effective in reducing the contribution to the TSI variation.  The extended pattern of smaller brighter and dimmer regions outside the active areas is a combination of contributions from the \Qone, \N\ and \Pz\ groups.

\section{The Solar Minimum Slope Difference}
\label{SMS}

We discussed above the inability of AutoClass-assigned areas to reproduce the long-term trend of decreasing TSI as measured by Virgo -- the Virgo TSI has a greater rate of decrease than is produced by solar surface magnetic features.  Our above conclusion could be a consequence of the tight constraints on coefficient values imposed by the large amplitude variations in the TSI due to the modulation of stationary plage and sunspot regions carried by rotation across the apparent solar surface.  By temporally smoothing the time series to eliminate the rotational modulation of the TSI record, the least square fitting may be freed to fit the long-term trend mismatch.  We have tested this supposition by carrying out a set of modified fits using both smoothed indices and the smoothed Virgo TSI data.  The degree of smoothing is defined by the Gaussian width parameter $\sigma$ which we have taken to be 3, 31 and 101 successive points.  Our points are separated by one or more days since observing conditions do not permit data acquisition every day so that the actual smoothing period varies but the number of observations smoothed does not.  After smoothing, the data sets were subsampled in order to retain only independent time sequence points and the regression fit using the bootstrap approach was repeated.  To determine whether these smoothed fits permit a better representation of the long-time trends, we define two temporal averaging periods: $\mathsf{Ave_{22m}}(X)$ and $\mathsf{Ave_{23m}}(X)$
which are 12-month averages of $X$ at the beginning of our series May, 1996 to April, 1997 and August, 2007 to July, 2008 respectively.  Recalling that $\delta TSI$ is the observed minus model, we define a solar minimum slope difference $\delta\mathsf{SMS}$ as:
\begin{eqnarray}
\delta\mathsf{SMS} &=& \left[\mathsf{Ave_{23{\rm m}}}(\delta TSI) - \mathsf{Ave_{22{\rm m}}}(\delta TSI)\right]/
(t_{23{\rm m}}-t_{22{\rm m}})\ .
\end{eqnarray}
For the Virgo observations the solar minimum slope is $-0.030$W$\,\!$m$^{-2}$yr$^{-1}$.
The smoothing and resampling reduces the number of constraints on the model so we considered three sets of class groups:
\begin{itemize}
\item[6 variables -- ] The class groups discussed above.
\item[7 variables -- ] The class groups discussed above except for the \Qz\ case which was split so that class C0 was one group and classes C4 and C6 were considered a second group.
\item[18 variables -- ]  All the original classes were considered as independent.
\end{itemize}

\begin{table}
\caption{Properties of fits}\label{tab3}
\begin{tabular}
[t] {@{}c@{}rr@{}r@{}rr@{\hfill}r@{}r@{}rr@{\hfill}r@{}r@{}r}
&&\multicolumn{3}{c}{$\delta\mathsf{SMS}$ (W$\,\!$m$^{-2}$yr$^{-1}$)}&
&\multicolumn{3}{c}{$r_P$ with trend}&
&\multicolumn{3}{c}{$r_P$ detrended}\\[1pt]
&&\multicolumn{3}{c}{$\sigma$ (days)}&&\multicolumn{3}{c}{$\sigma$ (days)}&
&\multicolumn{3}{c}{$\sigma$ (days)}\\[1pt]
\cline{3-5}\cline{7-9}\cline{11-13}
\vspace{-3pt}
\raise1pt\hbox{\strut}Vars.&
&\multicolumn{1}{c}{none\hspace*{-5pt}}&\multicolumn{1}{r}{3}&\multicolumn{1}{r}{31\hspace*{3pt}}&
&\multicolumn{1}{c}{none\hspace*{-5pt}}&\multicolumn{1}{r}{3}&\multicolumn{1}{r}{31\hspace*{3pt}}&
&\multicolumn{1}{c}{none\hspace*{-5pt}}&\multicolumn{1}{r}{3}&\multicolumn{1}{r}{31\hspace*{3pt}}\\[3pt]
\cline{1-1}\cline{3-5}\cline{7-9}\cline{11-13}

\raise1pt\hbox{\strut}6&&$-$0.025&~~$-$0.024& ~~$-$0.028&&0.888&~~0.926&~~0.909&&0.898&~~0.941&~~0.927\\
7&&$-$0.023& ~~$-$0.023&  ~~$-$0.026&&0.890&~~0.928&~~0.909&&0.897&0.940&~~0.926\\
18&&$-$0.023& ~~$-$0.022&  ~~$-$0.024&&0.953&~~0.958&~~0.925&&0.963&0.971&~~0.931\\[-3pt]
\hline
\end{tabular}
\end{table}

\begin{figure}
\begin{center}
\parbox{4.5in}{\begin{center}
\resizebox{4.5in}{!}{\includegraphics{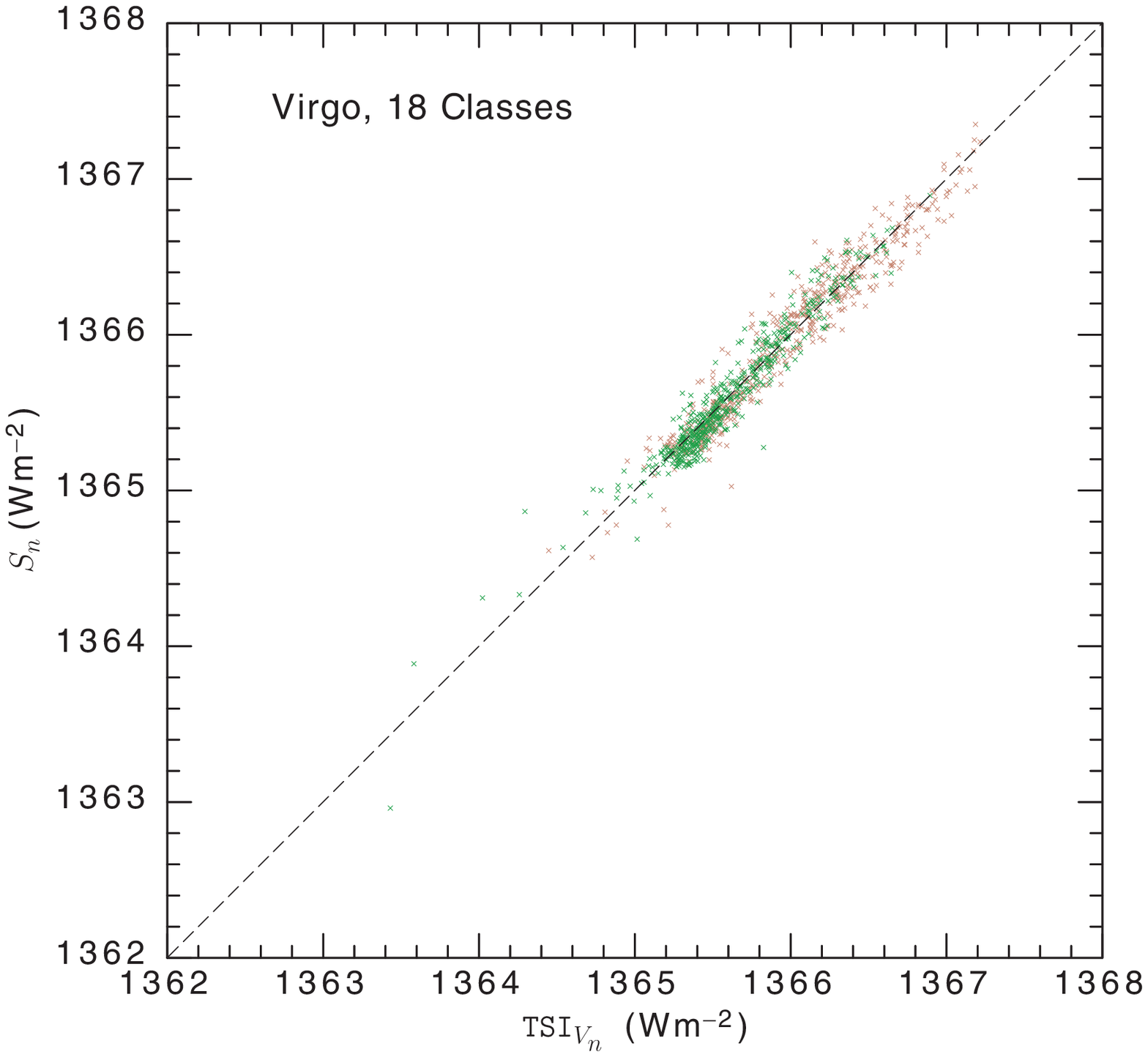}}
\end{center}
}
\caption{The relationship of modeled values of the TSI versus the 3-point smoothed and detrended observed values of the TSI during the period 1996 to 2008 using the same format as Figure~\ref{fig1}.}
\label{fig9}
\end{center}
\end{figure}

In Table~\ref{tab3} we report the values of $\delta\mathsf{SMS}$ for six cases defined by the above three numbers of variables and by the two smoothing range.  An 18-month averaging interval for the periods at the beginning and end of the series was tried and yield essentially the same results as given in Table~\ref{tab3}.  We find that both smoothing and reducing the number of fitting variables make $\delta\mathsf{SMS}$ more negative.  As an additional indication of the effect of the trend on our fits, we evaluate the Pearson correlation coefficient $r_P$ for these cases both with and without the trend between the two solar minima.  For the detrended cases, we adjust the observed values according to:
\begin{eqnarray}
\deltaTSI + 0.0254 \hbox{W$\,\!$m$^{-2}\,$yr$^{-1}$}(t-2002.0) &\longrightarrow& \delta TSI
\end{eqnarray}
where the notation indicates that the revised $\delta TSI$ is the result of the addition left of the arrow.
The improvement in the value of $r_P$ for the 3-point smoothed case is a result of the reduction of noise in the MWO model due to observational factors.  Our highest value of $r_P=0.971$ indicates a remarkable degree of success in the method's ability to reproduce the observed variations in the total solar irradiance.  The scatter diagram relationship between the model and the observations is shown in Figure~\ref{fig9}.

These results strengthen our earlier conclusion that the long-time trend in the difference between the model and the observations is not a consequence of a change in the magnetic coverage of the solar surface.  If this discrepancy were to be explained by changes in the solar radius and/or surface temperature it would require a decrease of about 80 milli arc-sec and/or 0.3$^\circ$K over the spanned period of 11.25 years. The observed radius change is opposite to what might be required to explain the slope excess or too small to influence our result \cite{2004ApJ...613.1241K,2008ApJ...676..651D,2008SoPh..247..225D}.  The temperature change is difficult to measure directly and is closely related to the TSI data such as we are discussing here \cite{2005MmSAI..76..961F}.  Since our analysis already includes the effect of the changes in surface coverage by different magnetic classes, this additional change must come from the temperature of one of the quiet classes -- what has been referred to by \inlinecite{2003SoPh..212..227L} as the immutable basal temperature.  The rate of change in this immutable basal temperature required by the $\delta\mathsf{SMS}$ is small compared to the amplitudes these authors discussed.

\section{Discussion}

The problem of identification and measurement of different parts of the solar surface along with the determination of their relationship, if any, with variability of the solar energy output has long been recognized and studied, see for example the reviews by \citeauthor{1988AdSpR...8..263L}\shortcite{1988AdSpR...8..263L,1997ARA&A..35...33L}.  The strength and geometry of magnetic fields is the underlying factor that distinguishes one portion of the solar surface from others.  Work by \inlinecite{1975ApJ...200..747S} showed that there is a direct relationship between the magnetic field strength and the strength of emission in the radiation near the core of the CaII K spectral line.  Spectroheliograms utilizing this spectral band have been used to identify facular and sunspot regions and estimate their impact on the TSI \cite{1978ApJ...226..679W,1984ApJ...282..776S,1998A&A...335..709F,1999ApJ...515..812H}.  The He I line at $\lambda10830$\AA\ has also been used to estimate the impact of faculae and network on the TSI \cite{1988ApJ...328..347F}.  Although the strength of the magnetic field is assumed to be the underlying cause, these papers have all used the proxy of the CaII K emission or the HeI $\lambda10830$\AA\ emission as an index to estimate the field strength.  The study by \inlinecite{2006A&A...460..583W} has used magnetic fields from the NSO-512 and the NSO-SPM at the KPVT to model the TSI variations in solar cycles 21 to 23.  The spatial resolution of these magnetograms is $1.14\arcsec\times 1.14\arcsec$ and the magnetic field measurements have a noise level of 7.5 G to 9 G.  This study also identified spots using continuum images.  For a period from 1992 to 2003 these authors obtained an optimized reconstruction of the TSI having $r_c=0.94$. 
The identification of features from their geometry and contiguity on the solar surface is a common goal of all these investigations.  While direct inspection is sometimes used, it is desireable to have a system that carries out this identification automatically so that potential investigator bias can be removed.  Two studies have implemented such automatic pattern recognition and classification systems \cite{2002ApJ...568..396T,2008SoPh..248..323J}.  The generally recognized features are sunspots, faculae and the network.  Of these the identification and measurement of the network has been most problematic.

Our approach differs from previous investigations in several ways:
\begin{enumerate}
\item The classification completely abandons all image-based information.  All pixels are treated equally.
\item We use a new type of data based on a line intensity ratio.  This quantity is sensitive to atmospheric properties not previously included -- the brightness in the upper photosphere/lower chromosphere.
\item Our magnetic field measurement has a noise level of about 0.2 G \cite{2009SoPh..255...53U} for the images used that are averages of 4 to 15 individual magnetograms.
\end{enumerate}
The resulting classification has allowed us to reproduce the observed TSI with $r_P$ reaching 0.97 for the case where the Virgo observations have the long-term trend removed and some of the MWO noise is reduced by smoothing with a 3-point wide Gaussian.  This high coefficient indicates that we need to take the classification seriously even though there are a number of deficiencies that are evident from the comparison to the SBI image.  The cancellation effect illustrated by the alternating sign of the coefficients for individual classes among the groups we have called \Pz\ and \Pone\ are indicative of an inadequacy of the classification -- probably a consequence of our low spatial resolution.  However, one strong aspect of our study is the low noise for the magnetic field strengths.  This is partly a consequence of our large size pixels.  If a similar level of noise could be achieved with pixels with sub-arcsecond size, the classification might be better.  Another way the classification could be improved is through the addition of more measured quantities such as a true continuum intensity that would give better identification of sunspot areas.  The MWO system does not provide such a quantity so its addition will require merging of an external data set wherein the temporal and spatial scales will need to be adjusted to match those of the MWO data set.  Other observational systems that are internally consistent do not provide a quantity having the properties of the $\Ir$ ratio we have used.  Part of the success of our analysis has come from the fact that this ratio is sensitive to the upper photosphere instead of the mid chromosphere.  However, it is possible that other combinations of quantities could be used in the AutoClass system with success similar to ours.

\section{Conclusions} 
      \label{S-conclusions}  
	This analysis holds out the possibility of creating an on-going, accurate, independent estimate of TSI variations from the MWO ground based observations which could be used to fill in gaps in the satellite record.  Comparison of the solar surface features seen in the TSI images created from the various TSI measurements may shed light on the sources of disagreement between various TSI measurements.  Further, the spatial resolution of these "images" should assist in identifying with greater accuracy the particular solar surface regions associated with TSI variations
and could help in addressing questions such as those raised by \inlinecite{2008SoPh..248....1F}.

\acknowledgements{This research has been supported over the years by a number of grants from NASA, NSF and ONR.  The most recent observations and analysis has been supported by NASA grant to Stanford 16165880-26967-G and NSF grant to UCLA NSF ATM-0517729.}





\end{article} 

\end{document}